\documentclass[preprint,tightlines,amsmath,amssymb,aps,prl,floatfix]{revtex4-2}
\usepackage{graphicx}
\usepackage{dcolumn}
\usepackage{bm}
\usepackage{hyperref}
\usepackage[mathlines]{lineno}
\usepackage[T1]{fontenc}
\begin{document}
\title{Large Deviations for the Spatial and Energy Distributions of Systems of Classical Ensemble}
\author{D. P. Shinde}
\email[]{dpshinde.physics@gmail.com}
\affiliation{168-A, Panchsheel Colony, Pachgaon, Kolhapur 416013, State-Maharashtra, India}
\date{\today}

\begin{abstract}
Boltzmann-Sanov and Cramer-Chernoff's theorems provide large deviation probabilities, entropy, and rate functions for the spatial distribution of systems and the total internal energy of an ensemble respectively. By the method of Lagrange's undetermined multipliers, the results of both theorems, the differentials of entropies or rate functions with respect to numbers of systems, establish the statistical equilibrium condition. We connect the large deviation statistics to the reversible and irreversible processes of the spatial arrangements of the systems and the energy exchange of the ensemble with the heat reservoir. We obtain the equalities between the differentials of entropies and other thermodynamic variables in the reversible processes and suggest inequalities of entropy productions for the irreversible processes. 
\end{abstract}
\maketitle
A cluster of phase points in the phase space of a given dimension, corresponding to various sets of the micro-states i.e. certain values of the generalized coordinates and momenta of the member systems, represents the micro-states of the whole ensemble \cite{LandauV5,Pathria11}. The macro-states and macroscopic variables, the measurable quantities of the systems, and the ensemble are defined for the various sets of the micro-states in their respective phase spaces \cite{LandauV5,Pathria11}. The statistical entropy, a state function of macroscopic variables and a measure of a multiplicity of micro-states for a given macro-state, is calculated for the spatial distribution of the systems over the states in the infinitesimal partitions of the phase space of an ensemble \cite{Planck91,Einstein90}. These infinitesimal partitions are called the elements of extension-in-phase or probability-region \cite{Gibbs02,Planck91,Einstein90}. The Gibbs probability-coefficient $P_{g}$, specified over either a multiple integral of micro-state variables or a simple integral of internal energy, defines a probability of finding a randomly selected system in the extension-in-phase \cite{Gibbs02,Planck91,Einstein90}. This definition, equivalently, is referred to as a ratio of the number of systems that fall within the extension-in-phase to the total number $\mathbb{N}$ of the systems \cite{Gibbs02}. For the systems of canonical and micro-canonical ensembles, $P_{g}$ or the density-in-phase $\mathcal{D}=\mathbb{N}P_{g}$ expresses, respectively, in terms of the exponential and delta functions of the internal energy \cite{LandauV5,Pathria11}. The canonical ensemble can be viewed as a finite or infinite set of micro-canonical systems and vice versa \cite{Gibbs02}. The temperature, pressure, or other external parameters determine the macro-states and alter the forms of $P_{g}$, the internal energy, and entropy of the systems and the ensemble \cite{LandauV5,Pathria11}.

Asymptotic mathematical methods \cite{Kolmogorov54,Khinchin49} are extremely useful for the calculations of the probabilities of macroscopic variables and entropies of the spatial arrangements of the systems and their energy exchange with the surroundings \cite{LandauV5,Pathria11}. In the former case, the whole ensemble is isolated from the surroundings whereas in the latter case, the ensemble is attached to the heat reservoir \cite{LandauV5,Pathria11,Hill03}. The laws of large numbers, central limit theorem (CLT), combinatorial counting method, the Stirling approximation, and the asymptotic approximation of the complex integral determine the probability distributions of the macroscopic variables, regarded as random variables, of the classical and quantum statistical mechanical systems \cite{LandauV5,Pathria11,Kolmogorov54,Khinchin49}. The probability distribution is related to the measure of the multiplicity of micro-states, and in turn to the entropy of a macro-state \cite{LandauV5,Pathria11,Hill03}. Clausius \cite{Clausius79} introduced the concept of entropy in his mechanical formulation of the second law of thermodynamics. For a statistical formulation of the second law, Boltzmann \cite{LandauV5,Pathria11,Boltzmann77} used the combinatorial counting method and the Stirling approximation for the factorials of large numbers to find first, the arbitrary spatial arrangements of systems and then, their most probable state with the maximum entropy. Besides, he suggested the origins of entropy productions in the irreversible processes \cite{Boltzmann77}. The steepest descent or saddle point method is used to calculate the mean value of the macroscopic variables using the asymptotic approximation of a complex integral \cite{LandauV5,Pathria11,Fowler22}. The methods of the most probable and mean value show, asymptotically and as a consequence of infinitely small fluctuations or deviations, identical results for the statistical equilibrium situation \cite{LandauV5,Pathria11}. These methods explore the normal deviations or fluctuations of the macroscopic random variables from their mean values \cite{LandauV5,Pathria11}. 

The  large deviation theory (LDT) \cite{Feller43,Linnik61,Sethuraman64,Petrov65,Varadhan66,Hoeffding67,Efron68,Ventsel70,Donsker75a,Donsker75b,Gartner77,Nagaev79,Ellis84,Bryc92} deals with  the asymptotic probabilities and their exponential convergence rates, and the mathematical properties of the rate functions of independent and dependent random variables beyond their normal deviations. The abstract formalism of the probability theory generalizes the large deviation bounds, properties of the rate function, and the form of the weak law of large numbers and convergence in probability of the random variables \cite{Varadhan84,Hollander00,Kobayashi11,Ellis06}. Moreover, in statistical mechanics \cite{Ellis06,Landford71,Oono89,Derrida98,Derrida02,Touchette09}, the LDT applied to the exclusion processes, many-particle systems, and chaotic dynamics.

This letter focuses on the calculations of the probabilities, entropies, and large deviation rate functions, using the simple mathematical forms of the Boltzmann-Sanov \cite{Boltzmann77,Sanov57} and Cramer-Chernoff's \cite{Cramer38,Chernoff52} theorems, respectively, for the spatial arrangements and a sum of internal energies of a large number of closed systems of a classical ensemble. In addition, for both the reversible and irreversible processes of the spatial arrangements and energy exchange, the findings of both theorems enable us to obtain the connections between the large deviation statistics and classical thermodynamics.  

We consider closed and isolated ensemble of the total volume $\mathbb{V}$ consisting of a large number $\mathbb{N}$ of macroscopic, distinguishable or indistinguishable, weakly interacting systems that share total internal energy $\mathbb{E}$ \cite{LandauV5,Pathria11}. No transport of matter takes place either in to or out of the systems or the closed ensemble during any process under consideration. The interaction energies between the systems are smaller than their internal energies. Therefore we have assumed that the probabilities of random variables of the quasi-independent systems do not affect each other \cite{LandauV5,Pathria11}. Following the mathematical theory of probability \cite{Varadhan84,Hollander00,Kobayashi11,Ellis06}, the phase space $\mathbb{R}$ of the ensemble is partitioned into $r$ partial regions or partitions $R_{1}, R_{2},.., R_{i},.., R_{r}$, where the subscript $i$ runs over all partitions \cite{LandauV5,Pathria11,Planck91,Einstein90}. Our first task is to calculate the joint probability distribution of the spatial arrangement of systems $(\{N_{i}\},(i=1,2,..,r))$ with internal energies $(\{E_{i}\}, (i=1,2,..,r))$ over the states of all $R_{i}$ partitions. Statistically \cite{Cramer62,Whittle00}, this can be viewed as a joint or multinomial probability mass distribution of all the events, the sets or subsets of the outcomes, of the generalized De Moivre-Bernoulli trials (See Supplementary Material \cite{Suppl} (SM \pageref{S1})). 

Let us see the approach in statistical mechanics for the spatial distribution of the systems \cite{LandauV5,Pathria11}. Here we first construct a multinomial spatial probability distribution and then calculate the Boltzmann-Sanov entropy and large deviation rate functions. Many distinguishable ways of a realization of the macroscopic configuration of the systems $(\{N_{i}\},(i=1,..,r))$ subject to two restrictive conditions are possible \cite{LandauV5,Pathria11}. These restrictive conditions, respectively, fix the total number $\mathbb{N}$ of the systems and total internal energy $\mathbb{E}$ of the ensemble, (i) $\sum^{r}_{i=1}N_{i}= \mathbb{N}$ and (ii) $\sum^{r}_{i=1}N_{i}E_{i}= \mathbb{E}= \mathbb{N}\overline{E}$, where $N_{i}\geq 0$, $E_{i}\geq 0$ \cite{LandauV5,Pathria11}. The ratio $\overline{E}=\mathbb{E}/\mathbb{N}$, the mean energy per system, remains finite as $\mathbb{E} \to \infty $ and $\mathbb{N} \to \infty $ \cite{LandauV5,Pathria11}. Since a system distributes randomly, and thus it may lie in any one of the states of an $i^{th}$ partition with a probability $G_{i}$. Then, the probability of exactly $N_{i}$ systems occupying the states in the same partition is $G_{i}^{N_{i}}$ \cite{LandauV5,Pathria11}. The multinomial spatial probability distribution $\mathbb{P}^{(b)}$ of the $\mathbb{N}$ quasi-independent and distinguishable systems, satisfying the two restrictive conditions, is obtained in such a way that exactly $N_{1}$ systems lie in states of $R_{1}$ partition, $N_{2}$ in the $R_{2}$,.., and exactly $N_{r}$ in the $R_{r}$. Using combinatorics, the probability distribution $\mathbb{P}^{(b)}$ takes the final form \cite{LandauV5,Pathria11,Cramer62,Whittle00,Hoeffding65,Kallenberg85} (SM \pageref{S2.1})
\begin{equation}
\mathbb{P}^{(b)}\left(\{N_{i}\}; i=1,..,r \right)  = \mathbb{N}! \prod^{r}_{i=1} \frac{G_{i}^{N_{i}}}{N_{i}!},\label{Eq1}
\end{equation}
where $G_{i}$'s are the relative volumes of the partitions, $0 \leq G_{i}\leq 1$ $(i=1,2,..,r) $, $\sum^{r}_{i=1}G_{i}=1$, and $\mathbb{N}! /(N_{1}!N_{2}!..N_{i}!...N_{r}!)$ is the multinomial coefficient \cite{Cramer62,Whittle00,Hoeffding65,Kallenberg85,Born49}. The natural logarithm on both sides of Eq. \eqref{Eq1} gives
\begin{eqnarray}
&&\log \mathbb{P}^{(b)}\left( . \right)  =\log \mathbb{N} ! + \sum_{i=1}^{r}\left(N_{i} \log G_{i}- \log N_{i}! \right) \label{Eq2}.\\
&& \log \mathbb{P}^{(b)}\left(. \right) = \sum_{i=1}^{r}N_{i}\log \frac{\mathbb{N}G_{i}} {N_{i}} + \mathcal{O}{(\log \mathbb{N})} \label{Eq3}.
\end{eqnarray}
Equation \eqref{Eq3} follows from the use of Stirling's approximation for the factorials of large numbers $\mathbb{N}$ and $N_{i}$ at Eq. \eqref{Eq2}, $\log N_{i} ! = N_{i} \log N_{i} - N_{i} +\mathcal{O}(\log \mathbb{N}_{i})$, and $\log \mathbb{N}! = \mathbb{N} \log \mathbb{N} - \mathbb{N}+\mathcal{O}(\log \mathbb{N})$ \cite{LandauV5,Pathria11,Ellis06}. The right-hand side of Eq. \eqref{Eq3} is the total Boltzmann-Sanov entropy $\mathbb{S}^{(b)}$ of the ensemble \cite{LandauV5,Pathria11,Ellis06}, and thus we write $k\log \mathbb{P}^{(b)}=\mathbb{S}^{(b)}+\mathcal{O}{(\log \mathbb{N})}$, where $k$ is the Boltzmann constant. The entropy $\mathbb{S}^{(b)}$ is the sum of the cross-entropy, $\mathbb{S}_{cross}= \sum_{i=1}^{r}N_{i}\log G_{i} $, and the entropy of mixing, $\mathbb{S}_{mix}=\sum_{i=1}^{r}N_{i}\log (\frac{\mathbb{N}}{N_{i}})$ \cite{Hanel14,Callen10,Haar66}. To obtain the Boltzmann-Sanov entropy and rate functions per system \cite{Varadhan84,Hollander00,Kobayashi11,Ellis06}, divide both sides of Eq. \eqref{Eq3} by $\mathbb{N}$, then we get
\begin{eqnarray}
\begin{split}
& \frac{1}{\mathbb{N}}\log \mathbb{P}^{(b)}\left(\{\frac{N_{i}}{\mathbb{N}}\}; i=1,..,r \right) = \sum_{i=1}^{r}F_{i}\log \frac{G_{i}} {F_{i}}+\tilde{\mathcal{O}}{(.)},\\ 
& \frac{1}{\mathbb{N}}\log \mathbb{P}^{(b)}(.)=S^{(b)}_{1}/k + \tilde{\mathcal{O}}{(.)} =  - I^{(b)}_{1} + \tilde{\mathcal{O}}{(.)},  
\end{split}
\label{Eq4}
\end{eqnarray} 
where $F_{i}= \frac{N_{i}}{\mathbb{N}}$ and $\sum_{i=1}^{r}F_{i}=1$ \cite{Kobayashi11,Ellis06}. From the right-hand side of the upper line of Eq. \eqref{Eq4}, we have a non-negative expression $\sum_{i=1}^{r}F_{i}\log(\frac{F_{i}}{G_{i}}) \geq 0$; the equality holds if and only if each $F_{i}$ equal to $G_{i}$ \cite{Kobayashi11,Ellis06}. The limit $\lim \limits^{}_{{\mathbb{N}} \to \infty}\log \mathbb{P}^{(b)}/\mathbb{N}$ on the left-hand side of Eq. \eqref{Eq4} exists, and as $\mathbb{N}\to \infty$, the term $\tilde{\mathcal{O}}{(\frac{\log \mathbb{N}}{\mathbb{N}})}$ converges to zero \cite{Kobayashi11,Ellis06}. On the right-hand side of Eq. \eqref{Eq4}, $S^{(b)}_{1} = k \sum_{i=1}^{r}F_{i}\log(\frac{G_{i}}{F_{i}})$, and  $I^{(b)}_{1}=-\frac{S^{(b)}_{1}}{k}=\sum_{i=1}^{r}F_{i}\log(\frac{F_{i}}{G_{i}})$ are, respectively, the Boltzmann-Sanov entropy \cite{Greiner95,Born49} and rate function \cite{Varadhan84,Hollander00,Kobayashi11,Ellis06} for a single system. In addition, $\mathbb{S}^{(b)} = \mathbb{N}S^{(b)}_{1}$ and $\mathbb{I}^{(b)}=\mathbb{N}I^{(b)}_{1}$ are corresponding functions for the whole ensemble. Therefore, the multinomial spatial probability distribution in Eq. \eqref{Eq1} becomes, $\mathbb{P}^{(b)} \approx e^{\frac{\mathbb{S}^{(b)}}{k}} = e^{-\mathbb{I}^{(b)}}=e^{-\mathbb{N}I^{(b)}_{1}}$. Furthermore, the rate function $I^{(b)}_{1}$ is also known as Kullback–Leibler (KL) divergence $D(F|G)$ between $F$ and $G$ \cite{Varadhan84,Hollander00,Kobayashi11,Ellis06,Kullback51,Cover06}. The supplementary material \cite{Suppl} (SM \pageref{S2.2}) provides the details of Eqs. \eqref{Eq1}, \eqref{Eq2}, \eqref{Eq3}, and \eqref{Eq4}, and findings in the case of quasi-independent and indistinguishable systems \cite{LandauV5,Pathria11,Zemansky17}. Also, we give results of the Boltzmann-Sanov entropy and the large deviation rate function for the multinomial spatial probability for the uniform distribution of quasi-independent and distinguishable systems.

Now we consider the probability distribution of the total internal energy $\mathbb{E}$ of the ensemble which exchanges only energy with the heat reservoir of temperature $T$. Our aim is to calculate the Cramer-Chernoff large deviation upper bound on $\mathbb{E}$ in terms of the generalized Bienayme-Chebyshev-Markov (BCM) probability inequality \cite{Cramer38,Chernoff52}. The real-valued random variable $\mathbb{E}$ is non-negative ($\geq 0$) and it can be written as either a sum of internal energies of weakly interacting individual systems, $\mathbb{E}= \sum_{l=1}^{\mathbb{N}} \varepsilon_{l}$ or a sum over all states of the partitions for which the systems have the same energies, $\mathbb{E}=\sum_{i=1}^{r}N_{i}E_{i}$ \cite{Pathria11,Kobayashi11}. The CLT states that the sum $\mathbb{E}$, under the generalized conditions, asymptotically, follows the Laplace-Gauss distribution \cite{LandauV5,Pathria11,Cramer62}. Moreover, the LDT deals with the asymptotic probabilities and rate functions of each $\epsilon_{l}$ and the sum $\mathbb{E}$ beyond their CLT scales \cite{Varadhan84,Hollander00,Kobayashi11,Ellis06}. 

For $\mathbb{E}$, let us define a new random variable $e^{-\beta \mathbb{E}}$ with a real-valued parameter $\beta$ ($=1/kT$) \cite{Kobayashi11,Ellis06}. Through the moment generating function (MGF), also known as a partition function, and its natural logarithm, a cumulant generating function (CGF), we can calculate, respectively, the moments and cumulants of $e^{-\beta \mathbb{E}}$ \cite{Kobayashi11,Ellis06}. Besides, the MGF or partition function $\mathbb{Z}^{(c)}(\beta)$ of the ensemble is defined as the Laplace transform of a structure-function \cite{Khinchin49} or the density of states $\Omega(\mathbb{E})$ \cite{Fowler80}, $\mathbb{Z}^{(c)}(\beta) = \mathcal{E}(e^{-\beta \mathbb{E}}) = \int^{\infty}_{0}e^{-\beta \mathbb{E}}\Omega(\mathbb{E})d\mathbb{E}$, where the symbol ${\mathcal{E}}$ denotes mathematical expectation \cite{Kobayashi11,Ellis06}. For the multinomial distribution of the quasi-independent and distinguishable \cite{LandauV5,Pathria11} or localized systems \cite{Fowler80}, the joint MGF or the partition function of the ensemble is equal to the $\mathbb{N}$-th power of the partition function $Z^{(c)}_{1}(\beta)$ of a single system ${\mathcal{E}}(e^{-\beta \sum_{i=1}^{r}N_{i}E_{i}}) = \sum^{}_{(\{N_{i}\}; 1\leq i \leq r)}\prod^{r}_{i=1} \mathbb{N}! \frac{(G_{i}e^{-\beta E_{i}})^{N_{i}}}{N_{i}!}= (\sum^{r}_{i=1}G_{i}e^{-\beta E_{i}})^{\mathbb{N}}= (Z^{(c)}_{1}(\beta))^{\mathbb{N}}$ \cite{LandauV5,Pathria11} (SM \pageref{S3.1}). Here, the sum in the parentheses is taken for all possible values of $N_{i}$ subject to two restrictive conditions, and the last two expressions follow the multinomial theorem \cite{Kobayashi11}. 

In the case of the quasi-independent and distinguishable systems, the Cramer-Chernoff  large deviation upper bound $\mathbb{P}^{(c)}$ \cite{Varadhan84,Hollander00,Kobayashi11,Ellis06,Cramer38,Chernoff52} (SM \pageref{S3.2}) on the probability inequality for $e^{-\beta \mathbb{E}}$ which exceeds the given threshold is obtained as 
\begin{equation}
\mathbb{P}^{(c)}(e^{-\beta \mathbb{E}} \geq e^{-\beta \mathbb{N}E}) \leq \frac{{\mathcal{E}}(e^{-\beta \mathbb{E}})}{e^{-\beta \mathbb{N}E}}= e^{\beta \mathbb{N}E}(Z^{(c)}_{1}(\beta))^{\mathbb{N}},\label{Eq5}
\end{equation}
where the inequality holds for $\beta \geq 0$ and $E$ is the internal energy of a system. The optimum value of $\beta$, chosen within the interval of convergence of the joint MGF, provides the tightest optimum upper bound \cite{Varadhan84,Hollander00,Kobayashi11,Ellis06}. To obtain the Gibbs canonical entropy \cite{LandauV5,Pathria11,Hill03} and Cramer-Chernoff large deviation rate function \cite{Varadhan84,Hollander00,Kobayashi11,Ellis06}, first, take the natural logarithm on both sides of Eq. \eqref{Eq5} and then divide both sides by $\mathbb{N}$. Finally, we have
\begin{eqnarray}
 \log \mathbb{P}^{(c)}(.)  && \leq \beta \mathbb{N}E + \mathbb{N}\log Z^{(c)}_{1}(\beta) \label{Eq6}. \\
\frac{1}{\mathbb{N}}\log \mathbb{P}^{(c)}(.) &&\leq S^{(c)}_{1}(E)/k = -I^{(c)}_{1}(E) \label{Eq7}.  
\end{eqnarray}
The limit $\lim \limits^{}_{{\mathbb{N}}\to\infty}\frac{\log \mathbb{P}^{(c)}}{\mathbb{N}}$ on the left-hand side of Eq. \eqref{Eq7} exists \cite{Varadhan84,Hollander00,Kobayashi11,Ellis06}. From the right-hand sides of Eqs. \eqref{Eq6} and \eqref{Eq7}, the functions $\frac{S^{(c)}_{1}(E)}{k}= \min \limits_{\beta \geq 0} (\beta E + \log Z^{(c)}_{1}(\beta))$ and $I^{(c)}_{1}(E)=-\frac{S^{(c)}_{1}(E)}{k}= \max \limits_{\beta \geq 0}(- \beta E - \log  Z^{(c)}_{1}(\beta))$ are, respectively, the Gibbs canonical entropy \cite{LandauV5,Pathria11,Hill03} and Cramer-Chernoff's rate function \cite{Varadhan84,Hollander00,Kobayashi11,Ellis06} for a single system. Moreover, for the whole ensemble, the respective functions are written as $\mathbb{S}^{(c)}(E) = \mathbb{N}S^{(c)}_{1}(E)$ and $\mathbb{I}^{(c)}(E) = \mathbb{N}I^{(c)}_{1}(E)$. Therefore, the probability inequality in Eq. \eqref{Eq5} takes the form $\mathbb{P}^{(c)} \leq e^{\frac{\mathbb{S}^{(c)}}{k}}= e^{-\mathbb{I}^{(c)}}=e^{-\mathbb{N}I^{(c)}_{1}}$. In general, both Gibbs canonical entropy and Cramer-Chernoff's rate function are the functions of the internal energy and external parameters, say, $\lambda_{n}(n=1,2,..)$ \cite{Khinchin49,Fowler80}. The minimum value of the parameter $\beta =\beta^{*}$, holding the external parameters $\lambda_{n}$ constant, gives the mean energy $\overline{E}= - (\partial \log Z^{(c)}_{1}(\beta)/\partial \beta)_{\beta =\beta^{*},\lambda_{n}}$ \cite{Khinchin49,Fowler80}. The rate function $I^{(c)}_{1}$ attains the minimum at mean energy, i.e, $I^{(c)}_{1}(\overline{E})=0$ \cite{Varadhan84,Hollander00,Kobayashi11,Ellis06}. 

We find that the Cramer-Chernoff's rate function is identical with the Gibbs index of probability $\eta$ (the natural logarithm of Gibbs probability-coefficient) \cite{Gibbs02}, $I^{(c)}_{1}(E)=\eta (E) = \beta (A-E)$, where $A(\beta)={-\beta}^{-1}\log Z^{(c)}_{1}(\beta)$ is the Helmholtz free energy \cite{LandauV5,Pathria11,Hill03} or the work-function \cite{Fowler80}. The Massieu potential \cite{Callen10} or Planck characteristic function \cite{Fowler80} per system is, $\psi_{m}(\beta) = -\beta A=\log Z^{(c)}_{1}(\beta)$. The functions $I^{(c)}_{1}(E)$ and $\log Z^{(c)}_{1}(\beta)$ or $\psi_{m}(\beta) $ show duality with each other via the Legendre-Fenchel transform \cite{Varadhan84,Hollander00,Kobayashi11,Ellis06}, $\log Z^{(c)}_{1}(\beta) =\psi_{m}(\beta) = \max \limits_{\beta} (-\beta E - I^{(c)}_{1}(E))$, and $I^{(c)}_{1}(E)=\max \limits_{E}(-\beta E-\psi_{m}(\beta))$. 

For the quasi-independent and indistinguishable \cite{LandauV5,Pathria11} or non-localized systems \cite{Fowler80}, we put the correct partition function of the ensemble $\mathcal{E}(e^{-\beta \mathbb{E}}) = \frac{Z^{(c)}_{1}(\beta))^{\mathbb{N}}}{\mathbb{N}!}$ \cite{LandauV5,Pathria11,Fowler80} in Eq \eqref{Eq5}, then take the natural logarithm on both sides of equation, and finally use the Stirling approximation for the factorial $\log \mathbb{N}! = \mathbb{N}\log \mathbb{N}- \mathbb{N}+\mathcal{O}{(\log \mathbb{N})}$. Putting the natural logarithm of the correct partition function $\log(\frac{eZ^{(c)}_{1}(\beta)}{\mathbb{N}})$ and following Eqs \eqref{Eq6} and Eq \eqref{Eq7}, we get the expressions for the Gibbs canonical entropy \cite{LandauV5,Pathria11,Fowler80}, Cramer-Chernoff's rate function, and the Helmholtz free energy (For details, see Supplementary Material \cite{Suppl} (SM \pageref{S3.2})). 
 
Let us show that the results of both theorems, the entropies or logarithms of the probability distributions, establish the statistical equilibrium condition. The total Boltzmann-Sanov and Gibbs canonical entropies without any additives, respectively, the right-hand sides of Eqs. \eqref{Eq3} and \eqref{Eq6}, are the functions of numbers $N_{i}$, $\mathbb{S}^{(b)} = k \sum_{i=1}^{r}\Phi^{(b)}(N_{i})$ and $\mathbb{S}^{(c)} = k \sum_{i=1}^{r}\Phi^{(c)}(N_{i})$, where $\Phi^{(b)}(N_{i}) = N_{i}\log (\frac{\mathbb{N}G_{i}}{N_{i}})$ and $\Phi^{(c)}(N_{i})=\beta N_{i} E_{i}+N_{i}\log Z^{(c)}_{1}(\beta)$ \cite{Greiner95,Born49}. Both the most probable and mean values of $N_{i}$ are the indicators of statistical equilibrium \cite{Greiner95,Born49}. The logarithms of probability distributions $\log \mathbb{P}^{(b)}$ and $\log \mathbb{P}^{(c)}$, or the entropies $\mathbb{S}^{(b)}$ and $\mathbb{S}^{(c)}$ reach their respective maximum values at the statistical equilibrium \cite{Greiner95,Born49}. The total differentials $d (\equiv \frac{\partial}{\partial N_{i}}dN_{i})$ of $\Phi^{(b)}(N_{i})$ and $\Phi^{(c)}(N_{i})$ with respect to the numbers $N_{i}$, keeping other parameters constant, are written as $ d(k\log \mathbb{P}^{(b)})= d\mathbb{S}^{(b)} \approx k \sum_{i=1}^{r}\log (\frac{\mathbb{N} G_{i}}{N_{i}}) dN_{i}$ and  $d(k\log \mathbb{P}^{(c)})=  d\mathbb{S}^{(c)}\approx k\sum_{i=1}^{r}[\beta E_{i} + \log Z^{(c)}_{1}(\beta)]dN_{i}$ respectively \cite{Greiner95,Born49}. 
In addition, the differentials $d(\log \mathbb{P}^{(b)})$ and $d(\log \mathbb{P}^{(c)})$, subject to the conditions $\sum_{i=1}^{r}dN_{i} =0$ and $ \sum_{i=1}^{r}E_{i}dN_{i} =0$, vanish at the equilibrium. Therefore, Lagrange's method of undetermined multipliers (SM \pageref{S4.1}) provides $\sum_{i=1}^{r}[\log (\frac{\mathbb{N} G_{i}}{N_{i}}) - \beta E_{i} - \log Z^{(c)}_{1}(\beta)] dN_{i} = 0$ \cite{Greiner95,Born49}. Thus, the equilibrium value $N_{i}^{*}$ of the numbers of systems is
\begin{equation}
\frac{N_{i}^{*}}{\mathbb{N}} = \frac{G_{i} e^{- \beta E_{i}}}{Z^{(c)}_{1}(\beta)} = G_{i} e^{\alpha - \beta E_{i}}\label{Eq8}, 
\end{equation}
where $e^{\alpha} = \frac{1}{Z^{(c)}_{1}(\beta)}=e^{\beta A}$ and the parameters $\alpha$ and $\beta$ are the Lagrange's undetermined multipliers \cite{Greiner95,Born49}. For the equal $G_{i}$, Eq. \eqref{Eq8} shows that the ratio $\frac{N_{i}^{*}}{\mathbb{N}}$ is equal to the probability $P(E_{i})= \frac{e^{-\beta E_{i}}}{Z^{(c)}_{1}(\beta)}$ of finding a system in a state of energy $E_{i}$ \cite{Greiner95,Born49}. The substitution of $N_{i}^{*}$ from Eq. \eqref{Eq8} into two restrictive conditions gives, respectively, the canonical partition function and the mean energy of a single system, $ Z^{(c)}_{1}(\beta) = \sum_{i=1}^{r}G_{i} e^{- \beta E_{i}}$ and $\overline{E}=\frac{\mathbb{E}}{\mathbb{N}}= - (\frac{\partial \log  Z^{(c)}_{1}(\beta)}{\partial \beta})_{\beta =\beta^{*}}$ \cite{Greiner95,Born49}. Putting this $\overline{E}$ in Eq. \eqref{Eq6}, we get the expression for the mean canonical entropy of a single system $S^{(c)}_{1}(\overline{E}) = -k\beta \frac{\partial \log  Z^{(c)}_{1}(\beta)}{\partial \beta}+k\log Z^{(c)}_{1}(\beta)$ \cite{Greiner95,Born49}. See Supplementary Material \cite{Suppl} (SM \pageref{S4.2}) for the statistical equilibrium condition, similar to Eq. \eqref{Eq8}, in the case of quasi-independent and indistinguishable systems.

Using Eqs. \eqref{Eq1} and \eqref{Eq5} and our notations, for a coupled model of an ensemble and a heat reservoir \cite{Fowler80}, we obtain the probability distribution of the macro-state of the systems $(\{N_{i}\},(i=1,2,..,r))$ \cite{Fowler80} and the relative probability of occurrence of the fluctuations \cite{Reif65} where either $\mathbb{S}^{(b)}= \mathbb{S}^{(c)}$ or $\mathbb{S}^{(b)} \neq \mathbb{S}^{(c)}$ (See Supplementary Material \cite{Suppl} (SM \pageref{S5})).

We now see the connection of the large deviation statistics with thermodynamical reversible and irreversible processes via the differentials of entropies or rate functions. Clausius \cite{Clausius79} gave equality (inequality) between the differential changes in the entropy and heat for the reversible (irreversible) processes. Moreover, he also showed that the uncompensated transformation \cite{Clausius79} or the entropy production \cite{Prigogine52} is zero and strictly positive ($>0$), respectively, in the reversible and irreversible processes.
From the classical thermodynamics \cite{Kawai07,Feng08,Still12,Horowitz09,Prigogine52}, we know that $d\mathbb{W}\geq \Delta \mathbb{A} $, $d\mathbb{W}_{rev}= \Delta \mathbb{A}$, and $d\mathbb{W}_{irrev} > d\mathbb{W}_{rev}$, where $\mathbb{W}$'s and $\Delta \mathbb{A}$ refer to work and the difference in Helmholtz free energy respectively. Then, we have the criterion of irreversibility \cite{Prigogine52} $d\mathbb{W}_{irrev}-d\mathbb{W}_{rev}= d\mathcal{L}> 0$, where $d\mathcal{L}$ is the uncompensated transformation (dissipated heat or work). In the present study, both the spatial arrangements of systems and the energy exchange between the ensemble and the heat reservoir contribute to entropy production \cite{Boltzmann77}.  In both processes, the uncompensated transformations are equal to the differences in the differentials of the respective entropies and heat $Td\mathbb{S}^{(b)}-d\mathbb{Q}^{(b)}=d\mathcal{L}^{(b)} \geq 0$ and $Td\mathbb{S}^{(c)}-d\mathbb{Q}^{(c)}=d\mathcal{L}^{(c)}\geq 0$; the equalities (inequalities) hold for the reversible (irreversible) processes \cite{Clausius79,Prigogine52}. The $d\mathbb{Q}$'s and $d\mathbb{W}$'s are the non-exact differentials \cite{Greiner95,Born49}.

Consider the reversible and irreversible processes of the spatial distribution of the systems. It is shown that the Kullback–Leibler (KL) divergence is equal to $\beta$ times the mean of dissipated work $\overline{\mathbb{W}}_{diss}$, \cite{Kawai07,Feng08,Still12,Horowitz09}. Besides, it is the difference between the irreversible and reversible works, $\overline{\mathbb{W}}_{diss} = \mathbb{W}_{irrev}-\mathbb{W}_{rev}\geq 0$ \cite{Kawai07,Feng08,Still12,Horowitz09}.  Here, we assert that the Boltzmann-Sanov rate function $\mathbb{I}^{(b)}$  or the KL divergence $D(F|G)$ between $F$ and $G$, analogous to the findings of \cite{Kawai07,Feng08,Still12,Horowitz09}, is equal to $\beta$ times the mean of dissipated work, $\mathbb{I}^{(b)}=  \mathbb{N} \sum_{i=1}^{r}\frac{N_{i}}{\mathbb{N}}\log (\frac{N_{i}}{\mathbb{N}G_{i}})=\beta \overline{\mathbb{W}}^{(b)}_{diss} = \beta(\mathbb{W}^{(b)}_{irrev}-\mathbb{W}^{(b)}_{rev})$. The $\mathbb{W}^{(b)}_{irrev}$ and $\mathbb{W}^{(b)}_{rev}$($=\Delta A^{(b)}$) are derived in terms of the chemical potentials, respectively, from the cross-entropy $\mathbb{S}_{cross}$ and the entropy of mixing $\mathbb{S}_{mix}$ \cite{Hanel14,Callen10,Shegelski86,Landsberg87,Kaplan06,Job06,Haar66,Baierlein99,Hill13}. Heat $\mathbb{Q}^{(b)}$ is produced during the rearrangements of systems over the states in the partitions \cite{Born49}. The differential change in the Boltzmann-Sanov entropy is given as $-\frac{d\mathbb{S}^{(b)}}{k}=d\mathbb{I}^{(b)}=\beta(d\mathbb{W}^{(b)}_{irrev}-\mathbb{W}^{(b)}_{rev})$. 

In the reversible process of energy exchange between the ensemble and the heat reservoir, following Eq. \eqref{Eq6}, the differential of total Gibbs canonical entropy gives the equality, $\frac{d\mathbb{S}^{(c)}}{k}= -d\mathbb{I}^{(c)}= \beta (d\overline{\mathbb{E}}-\overline{d\mathbb{W}}^{(c)}) $, where $d\overline{\mathbb{E}}$ and $\overline{d\mathbb{W}}^{(c)}= -\beta^{-1} \sum^{}_{n}(\frac{\partial \log \mathbb{Z}^{(c)}(\beta)}{\partial \lambda_{n}})d\lambda_{n}$ are the changes in the mean internal energy and mean work-done respectively \cite{Greiner95,Born49}. Moreover, we know the relation between the heat change and entropy in the reversible process, $\beta d\mathbb{Q}^{(c)}= \frac{d\mathbb{S}^{(c)}}{k}$ \cite{Greiner95,Born49}. Furthermore, if we set $d\mathbb{W}^{(c)}_{irrev} = -\mathbb{X}^{(c)}d\mathbb{V}$, and $d\mathbb{W}^{(c)}_{rev} = d\mathbb{A}^{(c)}$ then $d\mathbb{W}^{(c)}_{irrev}-d\mathbb{W}^{(c)}_{rev}=d\mathcal{L}^{(c)}\geq 0$, where $\mathbb{X}^{(c)}$ and $\mathbb{V}$ are pressure and volume respectively \cite{Prigogine52}.

The total entropy of the ensemble can vary either by entropy production due to the spatial arrangements of the systems in the interior of the ensemble, or entropy exchange to or from the heat reservoir (exterior of the ensemble) \cite{Prigogine52}. These two contributions, corresponding to the differentials of the total Boltzmann-Sanov and Gibbs canonical entropies, are written as $d\mathbb{S}^{(j)}= d\mathbb{S}^{(b)}+d\mathbb{S}^{(c)} \geq 0$, where the equality $d\mathbb{S}^{(j)} =0$ gives the relation, $-d\mathbb{S}^{(b)}= d\mathbb{S}^{(c)}$ for a reversible process and the inequality ($>$ sign) indicates irreversibility. We suggest two inequalities of the entropy productions, analogous to those mentioned in the classical thermodynamics \cite{Prigogine52}, respectively, in the processes of the spatial arrangements of systems (matter) in the interior of the ensemble and the energy exchange between the ensemble and the heat reservoir (external surroundings), (i) $d\mathbb{S}^{(j)}-d\mathbb{S}^{(c)}= d\mathbb{S}^{(b)} \geq 0$, and (ii) $d\mathbb{S}^{(j)}-d\mathbb{S}^{(b)}=d\mathbb{S}^{(c)} \geq 0$, where equal (greater than) sign refer to the reversible (irreversible) processes.

Summary - This letter describes the applications of large deviation theorems for the distribution of the numbers $N_{i}$ and sum $\mathbb{E}$ of internal energies of the quasi-independent and closed classical systems. In both distributions, the large deviation probabilities decay exponentially as the ensemble size increases, i.e. as $\mathbb{N}\to\infty$. We find that the rate functions, obtained from both theorems, are negative of their respective entropies. For both distinguishable and indistinguishable quasi-independent systems, we calculate only the large deviation upper bound on the probability of total internal energy $\mathbb{E}$ of the ensemble. We identify that Cramer-Chernoff's rate function is identical with the Gibbs index of probability and shows the duality with the Massieu potential or Planck's characteristic function via the Legendre-Fenchel transform. We have explored the large deviation statistics to obtain the statistical equilibrium condition, a probability distribution of the macro-state of the systems in a coupled model of the ensemble and heat reservoir, and the equalities between the differentials of entropies and other thermodynamical variables in the reversible processes of the spatial arrangements of systems and energy exchange between the ensemble and the heat reservoir. Also, we suggest inequalities of the entropy productions for the irreversible processes. The supplementary material \cite{Suppl} contains the necessary details of our results. 



{\bf Supplementary Material : “Large Deviations for the Spatial and Energy Distributions of Systems of a Classical Ensemble"}

\section{(I) Statistics of the Multinomial Probability Distribution}\label{S1}
The multinomial probability mass distribution is defined for all the events, the sets or subsets, of the outcomes of the generalized De Moivre-Bernoulli trials \cite{Cramer62,Whittle00}. For a single trial, “a single system can occupy any one of the states of an $i$-th partition" is an $i$-th event with a probability $G_{i}$. Therefore, in $\mathbb{N}$ independent trials, the event $i$ occurs exactly $N_{i}$ times with the probability $G_{i}^{N_{i}}$ \cite{Cramer62,Whittle00}. The multinomial probability distribution $\mathbb{P}^{(b)}$ of all the conditionally independent events, subject to the restriction $\sum^{r}_{i=1}N_{i}= \mathbb{N}$, is obtained in such a way that the first event occurs exactly $N_{1}$ times, the second in $N_{2}$ times,..., and the $r$-th event in exactly $N_{r}$ times \cite{Cramer62,Whittle00}. Therefore, $\mathbb{P}^{(b)}$ is a product of probabilities of the random variables $X_{i}$ which exactly takes $N_{i}$ values in all events, $\mathbb{P}^{(b)}\left(\{N_{i}\}; i=1,..,r \right)= \mathbb{N}!\prod_{i=1}^{r}P(X_{i} = N_{i})$ \cite{Cramer62,Whittle00}. 

In a spatial distribution model \cite{Cramer62,Whittle00}, the positions of systems in the partitions are considered as random variables $X_{i}$. Let $\mathcal{E}(X_{i}) = \mathbb{N}G_{i}$ and $\mathbb{D}(X_{i})=\sqrt{\mathbb{N}G_{i}(1-G_{i})} \approx \sqrt{\mathbb{N}}G_{i}$ are the means and standard deviations (s.d.) of the random variables $X_{i}$ respectively.  The real-valued random variables $Y_{i}$'s in their standardized forms $Y_{i}=(N_{i}-\mathbb{N}G_{i})/\sqrt{\mathbb{N}}G_{i}$ $(i=1,2,...,r)$ have zero means and unit standard deviations \cite{Cramer62,Whittle00}. The De Moivre-Laplace theorem \cite{Cramer62,Whittle00,Cramer38} states that the joint distribution function $\mathcal{F}(Y_{i})$ $(i=1,2,...,r)$ of $Y_{i}$'s, independent of $\mathbb{N}$, asymptotically, tends to the multivariate normal distribution $\Phi(Y_{i})$. Moreover, the normal deviations of $Y_{i}$'s are of the order of $\sqrt{\mathbb{N}}$. Furthermore, the joint moment generating function (MGF) or characteristic function (CF) of $Y_{i}$'s $(i=1,2,...,r)$  leads, asymptotically, to the MGF or CF of the multivariate normal distribution \cite{Cramer62,Whittle00}. The deviations of $Y_{i}$'s may vary beyond their normal distribution scale i.e. of the order of $\mathbb{N}^{\zeta}$ $(0< \zeta <1)$ \cite{Cramer38,Sanov57}. The large deviation theory (LDT) concerns the asymptotic behavior of the probabilities of the random variables $Y_{i}$'s, $P(Y_{i} \geq C \mathbb{N}^{\zeta})$, where $C>0$ or $P(Y_{i} \geq \gamma(\mathbb{N})$, where $\gamma(\mathbb{N})=\mathcal{O}(\mathbb{N}^{1/2})$ \cite{Cramer38,Sanov57}. 

\section{(II) Large Deviations for the Multinomial Spatial Distribution} \label{S2}
\subsection{(i) Quasi-independent and Distinguishable systems}\label{S2.1}
 In the spatial arrangements of the quasi-independent and distinguishable systems \cite{Boltzmann77,Sanov57,LandauV5,Pathria11}, the multinomial spatial probability distribution $\mathbb{P}^{(b)}$ is,
\begin{equation}
\mathbb{P}^{(b)}\left(\{N_{i}\}; i=1,..,r \right) = \mathbb{N}! \prod^{r}_{i=1} \frac{G_{i}^{N_{i}}}{N_{i}!}\label{Eq9},
\end{equation}
where $N_{i}\geq 0$, $\sum_{i=1}^{r}N_{i}=\mathbb{N}$, $ 0 \leq G_{i}\leq 1$ $(i=1,2,..,r) $, and $\sum^{r}_{i=1}G_{i}=1$ \cite{Boltzmann77,Sanov57,LandauV5,Pathria11}. 
Taking the natural logarithm on both sides of Eq. \eqref{Eq9}, we get
\begin{equation}
\log \mathbb{P}^{(b)}\left(\{N_{i}\}; i=1,..,r \right)  = \log\mathbb{N}! + \sum_{i=1}^{r}\left(N_{i}\log G_{i}- \log N_{i}!\right) \label{Eq10}. \\ 
\end{equation}
Using Stirling's approximation for the factorials of the large numbers $\mathbb{N}$ and $N_{i}$, $\mathbb{N}! = \mathbb{N} \log  \mathbb{N}-  \mathbb{N} + \mathcal{O}{(\log \mathbb{N})}$ and $N_{i}! = N_{i} \log N_{i}- N_{i} + \mathcal{O}{(\log N_{i})}$ \cite{Boltzmann77,Sanov57,LandauV5,Pathria11}, Eq. \eqref{Eq10} becomes,
\begin{eqnarray}
\begin{split}
\log \mathbb{P}^{(b)}\left(\{N_{i}\}; i=1,..,r \right) &= \mathbb{N} \log \mathbb{N} - \mathbb{N} + \sum_{i=1}^{r}\left(N_{i}\log G_{i}-N_{i}\log N_{i} +  N_{i}\right) +\mathcal{O}{(\log \mathbb{N})}\label{Eq11},\\ 
& = \left(\sum_{i=1}^{r}N_{i}\right) \log \mathbb{N}+ \sum_{i=1}^{r}\left(N_{i}\log G_{i}-N_{i}\log N_{i}\right) +\mathcal{O}{(\log \mathbb{N})},\\ 
& = \sum_{i=1}^{r}\left( N_{i}\log \frac{\mathbb{N}G_{i}} {N_{i}}\right) +\mathcal{O}{(\log \mathbb{N})}.
\end{split}
\end{eqnarray}
The total Boltzmann-Sanov entropy is proportional to the natural logarithm of the probability distribution, $\mathbb{S}^{(b)} = k \log \mathbb{P}^{(b)}$, where $k$ is the Boltzmann constant. 
The division on both sides of the last line of Eq. \eqref{Eq11} by $\mathbb{N}$ gives the equations for the Boltzmann-Sanov entropy and large deviation rate functions per system \cite{Varadhan84,Hollander00,Kobayashi11,Ellis06,Chernoff52,Hoeffding65,Kallenberg85}, 
\begin{eqnarray}
\begin{split}
\frac{1}{\mathbb{N}}\log \mathbb{P}^{(b)}\left(\{\frac{N_{i}}{\mathbb{N}}\}; i=1,..,r \right) & = \sum_{i=1}^{r}\left(\frac{N_{i}}{\mathbb{N}}\log \frac{\mathbb{N}G_{i}} {N_{i}}\right) +\tilde{\mathcal{O}}{\left(\frac{\log \mathbb{N}}{\mathbb{N}}\right)} \label{Eq12}, \\
& = \sum_{i=1}^{r}\left(F_{i}\log \frac{G_{i}} {F_{i}}\right) +\tilde{\mathcal{O}}{\left(\frac{\log \mathbb{N}}{\mathbb{N}}\right)},\\ 
& = S^{(b)}_{1}/k+\tilde{\mathcal{O}}{(.)}=-I^{(b)}_{1}+\tilde{\mathcal{O}}{(.)},
\end{split}
\end{eqnarray}
where $F_{i}=\frac{N_{i}}{N}$, and $\sum^{\mathbb{N}}_{i=1} F_{i}=1$ \cite{Varadhan84,Hollander00,Kobayashi11,Ellis06,Chernoff52,Hoeffding65,Kallenberg85}. As $\mathbb{N} \to \infty$, $\tilde{\mathcal{O}}{(\frac{\log \mathbb{N}}{\mathbb{N}})} \to 0 $ \cite{Varadhan84,Hollander00,Kobayashi11,Ellis06,Chernoff52,Hoeffding65,Kallenberg85}. For a single system, $S^{(b)}_{1} = \frac{\mathbb{S}^{(b)}}{\mathbb{N}} \approx k \sum_{i=1}^{r}F_{i}\log (\frac{G_{i}}{F_{i}}) $, and  $I^{(b)}_{1} =\frac{\mathbb{I}^{(b)}}{\mathbb{N}} \approx -\frac{S^{(b)}_{1}}{k}=\sum_{i=1}^{r}F_{i}\log (\frac{F_{i}}{G_{i}}) $ are, respectively, the Boltzmann-Sanov entropy and rate functions.  

\subsection{(ii) Uniform Multinomial Probability Distribution of Systems} \label{S2.2}
For the multinomial spatial probability of the uniformly distributed systems or the joint distribution of equally likely events with the probabilities, $G_{i}=\frac{1}{r}$,$(i=1,2,..r)$ for each $i$, and following Eq. \eqref{Eq9}, \eqref{Eq10}, \eqref{Eq11}, and \eqref{Eq12}, we obtain expressions, respectively, for the multinomial probability $\mathbb{P}^{(b)}_{u}$ and its natural logarithm $\log \mathbb{P}^{(b)}_{u}$, total Boltzmann-Sanov entropy $\frac{\mathbb{S}^{(b)}_{u}}{k}$ and rate functions $\mathbb{I}^{(b)}_{u}$ of the ensemble \cite{Hollander00,Ellis06,Einstein90}.
\begin{eqnarray}
&& \mathbb{P}^{(b)}_{u}\left(\{N_{i}\}; i=1,..,r \right) = \mathbb{N}! \prod^{r}_{i=1}\frac{r^{-\mathbb{N}}}{N_{i}!}\label{Eq13}.  \\
&& \log \mathbb{P}^{(b)}_{u}\left(\{N_{i}\}; i=1,..,r \right) = \log \mathbb{N}!- \sum_{i=1}^{r} [\mathbb{N} \log r -  \log N_{i}!], \\ \nonumber
&& \log \mathbb{P}^{(b)}_{u}\left(\{N_{i}\}; i=1,..,r \right)  \approx -\mathbb{N}\log r + \sum_{i=1}^{r}N_{i} \log \left(\frac{\mathbb{N}}{N_{i}}\right) = \frac{\mathbb{S}^{(b)}_{u}}{k} =- \mathbb{I}^{(b)}_{u} \label{Eq14}. \\
&& \frac{1}{\mathbb{N}}\log \mathbb{P}^{(b)}_{u}\left(\{\frac{N_{i}}{\mathbb{N}}\}; i=1,..,r \right) \approx -\log r + \sum_{i=1}^{r}F_{i}\log \left(\frac{1} {F_{i}}\right)=\frac{S^{(b)}_{1,u}}{k} = - I^{(b)}_{1,u}\label{Eq15}.
\end{eqnarray}
where $\frac{S^{(b)}_{1,u}}{k} = -\log r - \sum_{i=1}^{r}F_{i}\log F_{i} $, and $I^{(b)}_{1,u} = \log r + \sum_{i=1}^{r}F_{i}\log F_{i}$ are the entropy and rate function for a single system respectively \cite{Hollander00,Ellis06,Einstein90}. 

\section{(III) Large Deviations for the Total Internal Energy of the Ensemble} \label{S3}
\subsection{(i) Moment Generating Function or Partition Function of the Ensemble} \label{S3.1}
The generalized Bienayme-Chebyshev-Markov (BCM) probability inequality expresses in terms of the moment generating function (MGF) and a given threshold \cite{Cramer62,Whittle00,Kobayashi11,Ellis06}. For a multinomial distribution \cite{Greiner95}, the joint MGF or partition function of the whole ensemble, subject to the constant total number of systems and total internal energy $\mathbb{E}$ of the ensemble, is written as 
\begin{eqnarray}
{\mathcal{E}}(e^{-\beta \mathbb{E}})= {\mathcal{E}}(e^{-\beta \sum_{i=1}^{r}N_{i}E_{i}}) && = \prod^{r}_{i=1} \mathbb{N}! \frac{G_{i}^{N_{i}}}{N_{i}!} (e^{-\beta \sum_{i=1}^{r}N_{i}E_{i}}) = \sum^{}_{(\{N_{i}\}; 1\leq i \leq r)}\prod^{r}_{i=1} \mathbb{N}! \frac{(G_{i}e^{-\beta E_{i}})^{N_{i}}}{N_{i}!} \label{Eq16},\\ 
&& = \sum^{}_{(\{N_{i}\}; 1\leq i \leq r)}\frac{\mathbb{N}!}{N_{1}!N_{2}!....N_{r}!} [(G_{1}e^{-\beta E_{1}})^{N_{1}}*....*(G_{r}e^{-\beta E_{r}})^{N_{r}}]\label{Eq17},\\
&& = \left(\sum^{r}_{i=1}G_{i}e^{-\beta E_{i}}\right)^{\mathbb{N}} = \left(Z^{(c)}_{1}(\beta)\right)^{\mathbb{N}}\label{Eq18},
\end{eqnarray}
where the $\mathcal{E}$ symbol on the left-hand side of Eq. \eqref{Eq16} denotes the mathematical expectation \cite{Kobayashi11,Ellis06}. The right-hand side of Eq. \eqref{Eq18} follows the multinomial theorem \cite{Greiner95}. The term $Z^{(c)}_{1}(\beta)$ is the partition function of a single system. 
The MGF or partition function $\mathbb{Z}^{(c)}(\beta)$ of the ensemble is also defined as the Laplace transform of the density of states $\Omega(\mathbb{E})$\cite{Greiner95,Fowler80} or a structure-function \cite{Khinchin49}, $\mathbb{Z}^{(c)}(\beta) = \mathcal{E}(e^{-\beta \mathbb{E}}) = \int^{\infty}_{0}e^{-\beta \mathbb{E}}\Omega(\mathbb{E})d\mathbb{E}$.
For the quasi-independent systems and distinguishable systems, the partition function of the ensemble is $\mathbb{Z}^{(c)}(\beta)= \mathcal{E}(e^{-\beta \mathbb{E}})= \mathcal{E}(\prod^{\mathbb{N}}_{l=1}e^{-\beta \varepsilon_{l}}) = \prod^{\mathbb{N}}_{l=1}\mathcal{E}(e^{-\beta \varepsilon_{l}})=[\mathcal{E}(e^{-\beta \varepsilon_{l}})]^{\mathbb{N}} =[Z^{(c)}_{1}(\beta)]^{\mathbb{N}} $, where 
$\mathcal{E}(e^{-\beta \varepsilon_{l}}) =  \int^{\infty}_{0}e^{-\beta \varepsilon_{l}}\omega(\varepsilon_{l})$ is the Laplace transform of the density of states $\omega(\varepsilon_{l}) $ of a single system \cite{Greiner95,Khinchin49}. 

\subsection{(ii) Cramer-Chernoff Large Deviation Upper Bound}  \label{S3.2}
We define a probability inequality \cite{Varadhan84,Hollander00,Kobayashi11,Ellis06,Chernoff52} for the total internal energy of the ensemble, $\mathbb{P}^{(c)}( \mathbb{E} \leq \mathbb{N}E) = \mathbb{P}^{(c)}(-\mathbb{E} \geq -\mathbb{N}E) = \mathbb{P}^{(c)}(-\beta \mathbb{E} \geq - \beta\mathbb{N}E) =\mathbb{P}^{(c)}(e^{-\beta \mathbb{E}} \geq e^{-\beta \mathbb{N}E})$, where $\beta (= \frac{1}{kT})\geq 0$, $T$ is the absolute temperature, and $E$ is the energy of a single system. 

(a) For the quasi-independent and distinguishable \cite{Greiner95} or localized systems \cite{Fowler80}, applying the Bienayme-Chebyshev-Markov inequality to the exponential random variable $e^{-\beta \mathbb{E}}$, the Cramer-Chernoff large deviation upper bound $\mathbb{P}^{(c)}$\cite{Varadhan84,Hollander00,Kobayashi11,Ellis06,Chernoff52} is,
\begin{eqnarray}
\mathbb{P}^{(c)}(e^{-\beta \mathbb{E}} \geq e^{-\beta \mathbb{N}E}) && \leq \frac{{\mathcal{E}}(e^{-\beta \mathbb{E}})}{e^{-\beta \mathbb{N}E}} = \frac{{\mathcal{E}}(e^{-\beta \sum_{i=1}^{r}N_{i}E_{i}})}{e^{-\beta \mathbb{N}E}}= \frac{\prod^{r}_{i=1}{\mathcal{E}}(e^{-\beta N_{i}E_{i}})}{e^{-\beta \mathbb{N}E}} =e^{\beta \mathbb{N}E} \left(Z^{(c)}_{1}(\beta)\right)^{\mathbb{N}}
\label{Eq19},
\end{eqnarray}
where we have put MGF from Eq. \eqref{Eq18}. Take the natural logarithm on both sides of Eq. \eqref{Eq19} and then divide both sides by $\mathbb{N}$,
\begin{eqnarray}
\log \mathbb{P}^{(c)}(e^{-\beta \mathbb{E}} \geq e^{-\beta \mathbb{N}E}) &&  
\leq \mathbb{N} \left(\beta E + \log Z^{(c)}_{1}(\beta)\right) \label{Eq20}.\\
\frac{1}{\mathbb{N}} \log \mathbb{P}^{(c)}(e^{-\beta \mathbb{E}} \geq e^{-\beta \mathbb{N}E})&& \leq \beta E + \log Z^{(c)}_{1}(\beta)\label{Eq21}.
\end{eqnarray}

(b) In the case of quasi-independent and indistinguishable \cite{LandauV5,Pathria11,Greiner95} or non-localized systems \cite{Fowler80}, using the correct partition function $\mathcal{E}(e^{-\beta \mathbb{E}})=\mathbb{Z}^{(c)}(\beta)=\frac{Z^{(c)}_{1}(\beta))^{\mathbb{N}}}{\mathbb{N}!}$ \cite{LandauV5,Pathria11,Greiner95}, and following the steps of Eqs. \eqref{Eq19}, \eqref{Eq20}, and \eqref{Eq21}, we get the set of equations
\begin{eqnarray}
&& \mathcal{P}^{(c)}_{id}(e^{-\beta \mathbb{E}} \geq e^{-\beta \mathbb{N}E}) \leq \frac{(Z^{(c)}_{1}(\beta))^{\mathbb{N}}/\mathbb{N}!}{e^{-\beta \mathbb{N}E}}= e^{\beta \mathbb{N}E}\frac{(Z^{(c)}_{1}(\beta))^{\mathbb{N}}}{\mathbb{N}!} \label{Eq22}.\\
&& \log \mathcal{P}^{(c)}_{id}(e^{-\beta \mathbb{E}} \geq e^{-\beta \mathbb{N}E}) 
\leq  \mathbb{N}\beta E + \mathbb{N}\log Z^{(c)}_{1}(\beta) - \log \mathbb{N}! \label{Eq23}. \\
&&\frac{1}{\mathbb{N}} \log \mathcal{P}^{(c)}_{id}(e^{-\beta \mathbb{E}} \geq e^{-\beta \mathbb{N}E}\leq  \beta E + \log \left(\frac{eZ^{(c)}_{1}(\beta)}{\mathbb{N}}\right) \label{Eq24}.
\end{eqnarray}
We have used Stirling's approximation for $\mathbb{N}!$ in Eq. \eqref{Eq23}.

(c) For the quasi-independent and distinguishable or localized systems, the Gibbs canonical entropy $S^{(c)}_{1}(E)$ \cite{Gibbs02}, Cramer-Chernoff rate function $I^{(c)}_{1}(E)$ per system (See Eq. \eqref{Eq21}) \cite{Kobayashi11,Ellis06} are, respectively, written as $\frac{S^{(c)}_{1}(E)}{k} = \min^{}_{\beta \geq 0}(\beta E + \log Z^{(c)}_{1}(\beta))$ and $I^{(c)}_{1}(E) = \max^{}_{\beta \geq 0}(-\beta E - \log Z^{(c)}_{1}(\beta))= \max^{}_{\beta \geq 0}(\beta(A - E))$, where $A(\beta)={-\beta}^{-1}\log Z^{(c)}_{1}(\beta)$ is the Helmholtz free energy or the work function of a single system \cite{Greiner95,Fowler80}. 
In the case of quasi-independent and indistinguishable or non-localized systems, we put logarithm of the correct partition function $\log (\frac{eZ^{(c)}_{1}(\beta)}{\mathbb{N}})$ in the expressions of Gibbs canonical entropy (See Eq. \eqref{Eq24}), rate function, and Helmholtz free energy \cite{Greiner95,Fowler80}. Moreover, $\mathbb{S}^{(c)}(E) = \mathbb{N}S^{(c)}_{1}(E)$ and $\mathbb{I}^{(c)}(E) = \mathbb{N}I^{(c)}_{1}(E)$ are, respectively, the total Gibbs canonical entropy and rate function of the ensemble. 

We have found that the Cramer-Chernoff large deviation rate function is identical with the Gibbs index of probability $\eta$, (the natural logarithm of coefficient of probability $P_{g}$),  $I^{(c)}_{1}(E)= \beta(A-E) =\eta = \log P_{g}$  \cite{Gibbs02}. Thus we can write $P_{g} = e^{\eta}= e^{I^{(c)}_{1}(E)}$ \cite{Gibbs02}. Besides, the rate function, the negative of entropy, is also identical to the Boltzmann $H$-function \cite{Greiner95,Fowler80}. 

\section{(IV) Lagrange's Method of Undetermined Multipliers}\label{S4}
\subsection{(i) The Statistical Equilibrium Condition for Quasi-independent and Distinguishable systems}  \label{S4.1}
We define two functions of numbers $N_{j}$, $\Phi^{(b)}(N_{j})$ and $\Phi^{(c)}(N_{j})$ \cite{Greiner95,Born49} from Eqs. \eqref{Eq11} and \eqref{Eq20} respectively,
 \begin{eqnarray}
 \log \mathbb{P}^{(b)}\left(. \right) && = \sum_{j=1}^{r}\left(N_{j}\log \frac{\mathbb{N}G_{j}} {N_{j}}\right) +\mathcal{O}{(\log \mathbb{N})} = \sum_{j=1}^{r}\Phi^{(b)}(N_{j}) +\mathcal{O}{(\log \mathbb{N})}\label{Eq25}, \\
 \log \mathbb{P}^{(c)}(.) &&  = \sum_{j=1}^{r} \left(N_{j} \log Z^{(c)}_{1}(\beta)+\beta  N_{j}E_{j} \right)= \sum_{j=1}^{r}\Phi^{(c)}(N_{j}) \label{Eq26},
\end{eqnarray}
where we have used $\mathbb{N} =  \sum_{j=1}^{r} N_{j}$. From these Eqs. \eqref{Eq25} and \eqref{Eq26}, the respective total entropies are $\mathbb{S}^{(b)} = k\log \mathbb{P}^{(b)}\left(. \right) \approx k \sum_{j=1}^{r} N_{j}\log \frac{G_{j}} {N_{j}}+k \mathbb{N}\log \mathbb{N}$ and $\mathbb{S}^{(c)} = k\log \mathbb{P}^{(c)}\left(. \right) = k \sum_{j=1}^{r} \left(N_{j} \log Z^{(c)}_{1}(\beta)+\beta N_{j}E_{j}\right)$. 

The total differentials $d \equiv (\frac{\partial}{\partial N_{i}} dN_{i}) $ with respect to the numbers $N_{j}$, for $j=i$, on both sides of Eqs \eqref{Eq25} and \eqref{Eq26} provide, respectively, 
 \begin{eqnarray}
 d \log \mathbb{P}^{(b)}\left(. \right) && = \sum_{i=1}^{r}\frac{\partial \Phi^{(b)}(N_{i})}{\partial N_{i} }dN_{i}= \sum_{i=1}^{r}\left[\log \left(\frac{\mathbb{N}G_{i}}{N_{i}}\right)-\log e\right] dN_{i}\label{Eq27}, \\
d \log \mathbb{P}^{(c)}(.) &&  = \sum_{i=1}^{r}\frac{\partial \Phi^{(c)}(N_{i})}{\partial N_{i}}dN_{i} = \sum_{i=1}^{r}\left(\beta E_{i} + \log Z^{(c)}_{1}(\beta)\right) dN_{i} \label{Eq28}.
\end{eqnarray}
To obtain the statistical equilibrium condition, we use Lagrange's method of undetermined multipliers subject to conditions $\sum_{i=1}^{r}dN_{i} =0$ and $ \sum_{i=1}^{r}E_{i}dN_{i} =0$ \cite{Greiner95,Born49}. Using Eqs. \eqref{Eq27} and \eqref{Eq28}, and ignoring $\log e$, we get the equilibrium value of the numbers $N_{i}^{*}$,
\begin{eqnarray}
&& \sum_{i=1}^{r} \left[\log\left(\frac{\mathbb{N} G_{i}}{N_{i}}\right)- \beta E_{i} - \log Z^{(c)}_{1}(\beta)\right]dN_{i} = 0 \label{Eq29}.\\
&&\frac{N_{i}^{*}}{\mathbb{N}} = \frac{G_{i} e^{- \beta E_{i}}}{Z^{(c)}_{1}(\beta)} = G_{i} e^{\alpha - \beta E_{i}}\label{Eq30}.
\end{eqnarray}
Here $e^{\alpha} = \frac{1}{Z^{(c)}_{1}(\beta)} = e^{\beta A}$, and the parameters $\alpha$ and $\beta$ are Lagrange's undetermined multipliers \cite{Greiner95,Born49}. 

\subsection{(ii) The Statistical Equilibrium Condition for Quasi-independent and Indistinguishable systems} \label{S4.2}
The division by $\mathbb{N}!$ on the right-hand side of Eq. \eqref{Eq9} gives the multinomial spatial probability distribution $\mathcal{P}^{(c)}_{id}(.)$ for quasi-independent and indistinguishable systems \cite{LandauV5,Pathria11}. We obtain the equations for the logarithm of probability distribution and total Boltzmann-Sanov entropy respectively.
\begin{eqnarray}
&& \mathcal{P}^{(b)}_{id}\left(\{N_{i}\}; i=1,..,r \right) = \prod^{r}_{i=1} \frac{G_{i}^{N_{i}}}{N_{i}!} \label{Eq31} \\
&& \log \mathcal{P}^{(b)}_{id}(.)  \approx \sum_{i=1}^{r}\left(N_{i}\log \frac{G_{i}} {N_{i}} + N_{i} \right) \label{Eq32}.\\
&& \mathcal{S}^{(b)}_{id} = k \log \mathcal{P}^{(b)}_{id}(.) \approx k \sum_{i=1}^{r}\left(N_{i}\log \frac{G_{i}} {N_{i}}\right)+ k\mathbb{N} \label{Eq33}.
\end{eqnarray}
From Eq. \eqref{Eq22}, we get the logarithm of Cramer-Chernoff probability distribution and the Gibbs canonical entropy
\begin{eqnarray}
 \log \mathcal{P}^{(c)}_{id}(.) &&\approx \sum_{i=1}^{r} N_{i} \left(\log Z^{(c)}_{1}(\beta)+\beta E_{i}-\log \mathbb{N} +\log e \right) \label{Eq34}.\\
\mathcal{S}^{(c)}_{id} = k \log \mathcal{P}^{(c)}_{id}(.) && \approx k\sum_{i=1}^{r} \left(N_{i} \log Z^{(c)}_{1}(\beta)+\beta N_{i}E_{i} \right)+k\mathbb{N}-k\mathbb{N}\log \mathbb{N} \label{Eq35}.
\end{eqnarray}
Following Lagrange's undetermined multipliers method, we finally obtain
\begin{equation}
\frac{N_{i}^{*}}{\mathbb{N}} = \frac{G_{i} e^{- \beta E_{i}}}{Z^{(c)}_{1}(\beta)}\label{Eq36}.
\end{equation}
Equations \eqref{Eq30} and \eqref{Eq36} are identical. The differences between the entropies of indistinguishable and distinguishable cases of the systems are written as $\mathcal{S}^{(b)}_{id} - \mathbb{S}^{(b)}\approx k(\mathbb{N} -\mathbb{N}\log \mathbb{N})$ and $\mathcal{S}^{(c)}_{id} - \mathbb{S}^{(c)}\approx k(\mathbb{N}-\mathbb{N}\log \mathbb{N})$. 

Alternatively, we can obtain the equilibrium conditions \eqref{Eq30} and \eqref{Eq36}, by the same method of Lagrange, using the variations of the rate functions $I^{(b)}_{1}$ and $I^{(c)}_{1}$ with respect to $F_{i} (=\frac{N_{i}}{\mathbb{N}})$ respectively. 

\section{(V) Coupled Model of Ensemble and Heat Reservoir}\label{S5}
Fowler \cite{Fowler80} obtained a relation between probability and entropy in the case of a coupled model. In this model, the ensemble is immersed in a heat reservoir which has a very large number of localized systems than the ensemble. Using Eqs. \eqref{Eq9} and \eqref{Eq19}, the probability distribution $\mathbb{P}^{(f)}$ \cite{Fowler80} of the state of systems in the coupled model is, 
\begin{equation}
\mathbb{P}^{(f)} = \frac{\mathbb{P}^{(b)}}{\mathbb{P}^{(c)}}= \left(\mathbb{N}! \prod^{r}_{i=1}\frac{G_{i}^{N_{i}}}{ N_{i}!}\right).\left[\frac{e^{-\beta \mathbb{N}E}}{Z^{(c)}_{1}(\beta)^{\mathbb{N}}}\right] \label{Eq37}.
\end{equation}
The natural logarithm on both sides of Eq. \eqref{Eq37} supplies
\begin{equation}
\log \mathbb{P}^{(f)} =  \log \mathbb{N}! + \sum_{i=1}^{r} \left(N_{i} \log G_{i}-\log N_{i} ! \right) - \left[\beta \mathbb{N} E+ \mathbb{N}\log Z^{(c)}_{1}(\beta)\right] \label{Eq38}.
 \end{equation}
We can write the ratio of probabilities in Eq. \eqref{Eq37} as
\begin{eqnarray}
\begin{split}
& \mathbb{P}^{(f)} = \frac{\mathbb{P}^{(b)}}{\mathbb{P}_{max}^{(c)}}= e^{(\mathbb{S}^{(b)}-\mathbb{S}^{(c)}_{max})/k},\\ 
& \log \mathbb{P}^{(f)} = \log \left(\frac{\mathbb{P}^{(b)}}{\mathbb{P}_{max}^{(c)}} \right)=\frac{\mathbb{S}^{(b)}-\mathbb{S}^{(c)}_{max}}{k},
\end{split}
\label{Eq39}
\end{eqnarray}
where $\mathbb{S}^{(b)}= \mathbb{N} k \sum_{i=1}^{r} N_{i}\log(\frac{\mathbb{N}G_{i}}{N_{i}}) $ and $\mathbb{S}^{(c)}_{max}= \mathbb{N}\beta \overline{E}+ \mathbb{N}\log Z^{(c)}_{1}(\beta)$, respectively, the total Boltzmann-Sanov and Gibbs canonical entropies. In Eq. \eqref{Eq39}, $\mathbb{P}_{max}^{(c)}$ and $\mathbb{S}^{(c)}_{max})/k$ are the maximums, respectively, of Cramer-Chernoff's probability and Gibbs canonical entropy. Moreover, the Eq. \eqref{Eq39} can be seen as the relative probability of the occurrence of the fluctuations \cite{LandauV5,Pathria11,Reif65} where both these entropies may or may not be equal.  

For the quasi-independent and indistinguishable case, we have the similar expression
$\mathcal{P}^{(f)}_{id} = \frac{\mathcal{P}^{(b)}_{id}}{\mathcal{P}^{(c)}_{id}}= \left(\prod^{r}_{i=1}\frac{G_{i}^{N_{i}}}{ N_{i}!}\right).\left[\frac{\mathbb{N}! e^{-\beta \mathbb{N}\overline{E}}}{Z^{(c)}_{1}(\beta)^{\mathbb{N}}}\right]$.

\bibliography{JointRef} 

\end{document}